\documentclass[prd,aps,singlecolumn,preprintnumbers, showpacs, nofootinbib,superscriptaddress,notitlepage]{revtex4-1}

\usepackage{enumitem}
  \usepackage{color} 
  \usepackage{slashed}
  \usepackage{mathtools}
  \usepackage{amsmath}
  \usepackage{graphicx}
  \usepackage{bbold}
  \usepackage{amsfonts}
  \usepackage{fancyhdr}
  \usepackage{float}
  \usepackage{physics}
  \usepackage{caption}
  \usepackage{subcaption}
\newcommand{\Blue}[1]{\textcolor{black}{#1}}

\graphicspath{ {./images/} }

\begin{document}

\preprint{}

\title{Renormalon Effects in Quasi Parton Distributions}
\author{Wei-Yang Liu}
\email{r07222052@g.ntu.edu.tw}
\affiliation{Department of Physics, Center for Theoretical Physics, and Leung Center for Cosmology and Particle Astrophysics, National Taiwan University, Taipei, Taiwan 106}

\author{Jiunn-Wei Chen}
\email{jwc@phys.ntu.edu.tw}
\affiliation{Department of Physics, Center for Theoretical Physics, and Leung Center for Cosmology and Particle Astrophysics, National Taiwan University, Taipei, Taiwan 106}
\affiliation{Physics Division, National Center for Theoretical Sciences, National Taiwan University, Taipei 10617, Taiwan}


\begin{abstract}

\Blue{We investigate the renormalon ambiguity from bubble-chain diagrams in the isovector unpolarized quasi- parton distribution function (PDF) of a hadron. We confirm the assertion by Braun, Vladimirov and Zhang \cite{Braun:2018brg} that the leading IR renormalon ambiguity is an $\mathcal{O}(\Lambda^2_{\text{QCD}}/x^{2}P_z^2)$ effect, with $x$ the parton momentum fraction and $P_z$ the hadron momentum, together with a new $\mathcal{O}(\delta(x)\Lambda^2_{\text{QCD}}/P_z^2)$ contribution such that the quark number is conserved. This implies the convergence of the perturbative matching kernel between a quasi-PDF and a PDF would eventually fail for small $x$. However, in both the R-scheme designed to cancel the leading IR renormalon and 
the typically used RI/MOM scheme in lattice QCD for the same quasi-PDF, we find good convergence in the kernel based on three-loop bubble-chain diagram analyses. These results are encouraging for the quasi-PDF program. However, firm conclusions can only be drawn after the complete higher loop QCD calculations are carried out.}

\end{abstract}
\maketitle

\section{Introduction}

Large momentum effective theory (LaMET) enables computations of parton distributions of hadrons on a Euclidean lattice. LaMET relates equal-time spatial correlators (whose Fourier transforms are called quasi-distributions) to lightcone distributions in the infinite hadron momentum limit~\cite{Ji:2013dva,Ji:2014gla}. For large but finite momenta accessible on a realistic lattice, LaMET relates quasi-distributions to physical ones through a factorization theorem, which involves a matching coefficient and power corrections that are suppressed by the hadron momentum. The proof of factorization was developed in  Refs.~\cite{Ma:2017pxb,Izubuchi:2018srq,Liu:2019urm}.

Since LaMET was proposed, a lot of progress has been made in the theoretical understanding of the formalism~\cite{Xiong:2013bka,Ji:2015jwa,Ji:2015qla,Xiong:2015nua,Ji:2017rah,Monahan:2017hpu,Stewart:2017tvs,Constantinou:2017sej,Green:2017xeu,Izubuchi:2018srq,Xiong:2017jtn,Wang:2017qyg,Wang:2017eel,Xu:2018mpf,Chen:2016utp,Zhang:2017bzy,Ishikawa:2016znu,Chen:2016fxx,Ji:2017oey,Ishikawa:2017faj,Chen:2017mzz,Alexandrou:2017huk,Constantinou:2017sej,Green:2017xeu,Chen:2017mzz,Chen:2017mie,Lin:2017ani,Chen:2017lnm,Li:2016amo,Monahan:2016bvm,Radyushkin:2016hsy,Rossi:2017muf,Carlson:2017gpk,Ji:2017rah,Briceno:2018lfj,Hobbs:2017xtq,Jia:2017uul,Xu:2018eii,Jia:2018qee,Spanoudes:2018zya,Rossi:2018zkn,Liu:2018uuj,Ji:2018waw,Bhattacharya:2018zxi,Radyushkin:2018nbf,Zhang:2018diq,Li:2018tpe,Braun:2018brg,Detmold:2019ghl,Sufian:2020vzb,Shugert:2020tgq,Green:2020xco,Braun:2020ymy,Lin:2020ijm,Bhat:2020ktg,Chen:2020arf,Ji:2020baz,Chen:2020iqi,Chen:2020ody,Alexandrou:2020tqq,Fan:2020nzz,Ji:2020brr}. The method has been applied in lattice calculations of parton distribution functions (PDF's) for the nucleon~\cite{Lin:2014zya,Chen:2016utp,Lin:2017ani,Alexandrou:2015rja,Alexandrou:2016jqi,Alexandrou:2017huk,Chen:2017mzz,Lin:2018pvv,Alexandrou:2018pbm,Chen:2018xof,Alexandrou:2018eet,Lin:2018qky,Fan:2018dxu,Liu:2018hxv,Wang:2019tgg,Lin:2019ocg,Liu:2020okp,Lin:2019ocg,Zhang:2019qiq}, $\pi$~\cite{Chen:2018fwa,Izubuchi:2019lyk,Gao:2020ito} and $K$~\cite{Lin:2020ssv}  mesons. Despite limited volumes and relatively coarse lattice spacings, the state-of-the-art nucleon isovector quark PDF's, determined from lattice data at the physical point, have shown reasonable agreement~\cite{Lin:2018pvv,Alexandrou:2018pbm} with phenomenological results extracted from the experimental data. Encouraged by this success, LaMET has also been applied to $\Delta^{+}$~\cite{Chai:2020nxw} and twist-three PDF's~\cite{Bhattacharya:2020cen,Bhattacharya:2020xlt,Bhattacharya:2020jfj}, as well as gluon \cite{Fan:2020cpa}, strange and charm distributions~\cite{Zhang:2020dkn}. It was also applied to meson distribution amplitudes~\cite{Zhang:2017bzy,Chen:2017gck,Zhang:2020gaj} and
generalized parton distributions (GPD's)~\cite{Chen:2019lcm,Alexandrou:2020zbe,Lin:2020rxa,Alexandrou:2019lfo}. More recently, attempts have been made to generalize LaMET to transverse momentum dependent (TMD) PDF's~\cite{Ji:2014hxa,Ji:2018hvs,Ebert:2018gzl,Ebert:2019okf,Ebert:2019tvc,Ji:2019sxk,Ji:2019ewn,Ebert:2020gxr} to calculate the nonperturbative Collins-Soper evolution kernel~\cite{Ebert:2018gzl,Shanahan:2019zcq,Shanahan:2020zxr} and soft functions \cite{Zhang:2020dbb} on the lattice.
LaMET also brought renewed interests in earlier approaches~\cite{Liu:1993cv,Detmold:2005gg,Braun:2007wv,Bali:2017gfr,Bali:2018spj,Detmold:2018kwu,Liang:2019frk} and inspired new ones~\cite{Ma:2014jla,Ma:2014jga,Chambers:2017dov,Radyushkin:2017cyf,Orginos:2017kos,Radyushkin:2017lvu,Radyushkin:2018cvn,Zhang:2018ggy,Karpie:2018zaz,Joo:2019jct,Radyushkin:2019owq,Joo:2019bzr,Balitsky:2019krf,Radyushkin:2019mye,Joo:2020spy,Can:2020sxc}. For recent reviews, see, e.g., Refs. \cite{Lin:2017snn,Cichy:2018mum,Zhao:2020vll,Ji:2020ect,Ji:2020byp}. 

\Blue{
The quark PDF in a proton is related to a lightcone correlator
\begin{equation}
\label{eq:def_PDF}
Q(x)=\int_{-\infty}^{\infty}\frac{d\xi^{-}}{4\pi}e^{-ix\xi^{-}P^{+}}\bra{P}\bar{\psi}(\xi^{-})\gamma^+\text{exp}\left(-ig\int_0^{\xi^{-}} d\eta^{-}A^{+}(\eta^{-})\right)\psi(0)\ket{P} ,
\end{equation}
where the proton momentum $P^{\mu}$ is alone the $z$-direction, $P^{\mu}=(P^{0}, 0,0,P^{z})$, and $\xi^{\pm}=(t\pm z)/\sqrt 2$ is the lightcone coordinates. $x$ is the parton momentum fractions relative to the proton. $A^{+}$ is a gauge field and $\psi$ is a quark field. 
The expression in Eq.(\ref{eq:def_PDF}) is boost invariant alone the $z$-direction. 
The corresponding quasi-PDF is related to an equal time correlator~\cite{Ji:2013dva,Ji:2014gla}
\begin{equation}
\label{eq:def_quasi-PDF}
\tilde{Q}(x, P_z)=\int_{-\infty}^{\infty}\frac{dz}{4\pi}e^{ixzP^{}z}\bra{P}\bar{\psi}(z)\gamma^z\text{exp}\left(-ig\int_0^z dz'A^{z}(z')\right)\psi(0)\ket{P} .
\end{equation}
%
%
}
LaMET relates the flavor non-singlet quark quasi-PDF and PDF of a hadron through the factorization theorem
\begin{equation}\label{eq:matching}
\tilde{Q}(x, P_z, \Lambda')=\int_{-1}^1 \frac{dy}{|y|} Z(\frac{x}{y},y P_z,\Lambda',\Lambda) Q(y, \Lambda) +\mathcal{O}(\frac{1}{P_z^2}),
\end{equation}
where $\Lambda'$ and $\Lambda$ are renormalization parameters and PDF is defined in the infinite momentum frame with $P_z \to \infty$~\cite{Ji:2013dva,Ji:2014gla}. The (quasi-)PDF for negative momentum fraction corresponds to the (quasi-)PDF of the anti-particle. In particular,  
$Q(-|y|, \Lambda)=-\overline{Q}(|y|, \Lambda)$ with $\overline{Q}$ denoting an antiquark PDF.

Let us first discuss the case that both the quasi-PDF and PDF are defined in cut-off schemes with UV cut-off $\Lambda'$ and $\Lambda$, respectively. The factorization theorem is based on a large momentum expansion in powers of $1/P_{z}$. 
The $\mathcal{O}(1/P_z^2)$ correction has a similar convolutional structure to the first term: $\sum_{n=4,...}Z_{n}(\Lambda',\Lambda)\otimes Q_{n}(\Lambda)/P_{z}^{n-2}$. The 
scales follow the hierarchy $\Lambda' \gg P_z \gg \Lambda \gg \Lambda_{QCD}$, such that the short distance (compared with $1/\Lambda$) physics is encoded in the matching kernels (or Wilson coefficients) $Z$ and $Z_{n}$ while the long distance physics is encoded in the matrix elements $Q$ and $Q_{n}$ with $Q$ of twist-2 and $Q_{n}$ of twist-$n$. In this series, the long and short distance physics are strictly separated by the scale $\Lambda$. 

However, the PDF's extracted from experimental data are typically defined in the modified minimal subtraction ($\overline{\text{MS}}$) scheme. This is because the PDF's convolute with Wilson coefficients to form observables in high energy experiments,  and  it is technically much easier to compute the Wilson coefficients to higher loops in $\overline{\text{MS}}$ than a hard cut-off scheme. However,  there is no strict separation between long and short distance physics in the $\overline{\text{MS}}$ scheme. This means the Wilson coefficients could still contain non-perturbative physics called renormalons~\cite{Gross:1974jv, Lautrup:1977hs,tHooft:1977xjm} to cause slow convergence in their perturbative expansions. 

Mathematically, the renormalon effect could arise when Borel transform is applied to improve the convergence of a perturbation series. Then an inverse Borel transform is applied after the Borel series is summed. However, sometimes the bad convergence of the series can not be tamed by the Borel transform but appears as ambiguities in the inverse Borel transform. 
When $Q$ is defined in the $\overline{\text{MS}}$ scheme with renormalization scale $\mu$ and the quasi-PDF $\tilde{Q}$ is defined in the regularization-independent momentum subtraction (RI/MOM) scheme \cite{Stewart:2017tvs} with  scale $\mu'$, 
Eq.(\ref{eq:matching}) can be rewritten as
\begin{equation}\label{eq:matchingX}
\tilde{Q}(x, P_z, \mu')=\int_{-1}^1 \frac{dy}{|y|} Z(\frac{x}{y},y P_z,\mu',\mu) Q(y, \mu) +\mathcal{O}(\frac{1}{P_z^2}) \ .
\end{equation}
Since scale separation is not complete in $\overline{\text{MS}}$ and RI/MOM, low energy non-perturbative physics could still exist in $Z$, which slows down the convergence of the perturbative computation of $Z$. It can be shown that by resumming a class of bubble-chain diagrams in $Z$ with the help of Borel transform, there are ambiguities in the inverse Borel transform which behave like non-perturbative functions $Q_{n}/P_{z}^{n-2}$ with $n\ge4$. 

Since these ambiguities are associated with different choices of the integration paths in the inverse Borel transform, they are not physical.
They should be cancelled by the $Q_{n}$ terms. Therefore all the $Q_{n}$'s need to be defined consistently. An independent determination of a higher twist $Q_{n}$ is not meaningful unless the treatment of renormalon ambiguity in the lower twist $Q_{n}$ is specified. Furthermore, one can estimate the size of $Q_{n}$ by the size of the renormalon ambiguity in $Z$. It was through this analysis that Braun, Vladimirov and Zhang asserted that the leading IR renormalon ambiguity was an \Blue{$\mathcal{O}(\Lambda^2_{\text{QCD}}/x^{2}P_z^2)$ effect}~\cite{Braun:2018brg}, which was larger than the previous estimation $\mathcal{O}(\Lambda^2_{\text{QCD}}/P_z^2)$ based on dimensional analysis by a factor $1/x^{2}$. Therefore, this result has a big impact to the LaMET program. This motivates us to re-investigate the renormalon problem in quasi-PDF. 



\section{Renormalon Ambiguity in Quasi-PDF's}

In this section we first briefly review the Borel transform and the possible ambiguity in its inverse transformation. \Blue{After the discoveries of renormalons~\cite{Gross:1974jv, Lautrup:1977hs,tHooft:1977xjm} and further conceptual developments~\cite{tHooft:1977xjm,Parisi:1978bj,Parisi:1978az,David:1983gz,David:1985xj,Mueller:1984vh}, higher order behaviors of perturbation series in the context of renormalon ambiguities became relevant phenomenolically~\cite{Brown:1991ic,Zakharov:1992bx,Mueller:1992xz}.
(See, e.g., Ref.\cite{Beneke:1998ui} for a review).}
We will compute the renormalon ambiguity caused by the bubble-chain diagrams to the quasi-PDF. We will compute the Feynman diagrams in momentum space then compare our result with the coordinate space computation obtained in
Ref.~\cite{Braun:2018brg}.

We start with writing the matching kernel of Eq.(\ref{eq:matchingX}) as a series expansion in the strong coupling constant $\alpha_{s}$,
\begin{equation}
Z(\xi)=\delta(1-\xi)+\sum_{n=1}^{\infty}\left(\frac{\alpha_s}{4\pi}\right)^nz_n(\xi) \ ,
\label{eq:matching kernel}
\end{equation}
where we have only kept the $\xi=x/y$ dependence and dropped the other quantities for brevity. In the $\overline{\text{MS}}$ scheme with renormalization scale $\mu$, 
\begin{equation}\label{eq:running coupling}
\alpha_s(\mu)=\frac{4\pi}{\beta_0\ln\frac{\mu^2}{\Lambda_{\text{QCD}}^2}} \ ,
\end{equation}
where $\beta_0 \equiv \frac{11}{3}N_c-\frac{2}{3}n_f$, $N_c$ being the number of colors, and $n_f$ being the number of quark flavors, and
$\Lambda_{\text{QCD}}$ is the strong interaction scale.  
The series is more convergent after the Borel transform 
\begin{equation}\label{eq:Borel series}
B[Z(\xi)]=\beta_0 \delta(1-\xi)\delta(w)+\sum_{n=0}^\infty \left(\frac{w}{\beta_0}\right)^n\frac{z_{n+1}(\xi)}{n!} \ .
\end{equation}
After the Borel series is summed, one can perform an inverse Borel transform back to the original series: 
\begin{equation}
Z(\xi)=
\frac{1}{\beta_0}\int_0^{\infty} dw e^{-\frac{4\pi w}{\beta_0\alpha_s}}B[Z(\xi)] \ .
\end{equation}
If the integrant has poles on the real positive $w$ axis, then the above integral will depend on the path of integration on the complex $w$ plane. This uncertainty is called renormalon ambiguity. In an operator product expansion with the $\overline{\text{MS}}$ scheme, the renormalon ambiguities in the lower order Wilson coefficients will be cancelled by the higher order matrix elements. 

\Blue{The existence of the renormalon ambiguity is usually demonstrated in the large $n_{f}$ limit with powers of $(\alpha_{s } n_{f})$ counted as $\mathcal{O}(1)$ and summed to all orders. Then renormalon ambiguity can arise in the $\alpha_{s }$ expansion of these diagrams. 
However, it is also well known that QCD is not asymptotically free in the large $n_{f}$ limit. Therefore, even though the finite $n_{f}$ contribution can be formally included by an $1/n_{f}$ expansion, this expansion is not convergent in QCD. 
%
%
However, if we take the QCD $\beta$ function as a reference, it is probably reasonable to expect that subleading
$1/n_{f}$ corrections are about the same magnitude as the leading order. If this expectation turns out to be true, then the large $n_{f}$ analysis will still be useful.
Therefore, we focus on the leading diagrams in the large $n_{f}$ limit to study the renormalon ambiguity in the matching formula  between a quasi-PDF and a PDF defined in Eq.(\ref{eq:matchingX}). 
} 

%
%
 %
%
%

\Blue{When $(\alpha_{s } n_{f})$ is counted as $\mathcal{O}(1)$, the bubble chain diagrams that dress the gluon propagator in the second row of Fig.~\ref{fig:1loopR} are all $\mathcal{O}(1)$. The gray bubble, which is the gluon vacuum polarization, becomes a fermion loop diagram in the large $n_{f}$ limit. We will add the gluon loop contribution to the gluon vacuum polarization as an $1/n_{f}$ correction as performed in \cite{Beneke:1998ui,Braun:2018brg}. This is formally a subleading effect in the 
$1/n_{f}$ expansion, but numerically important in QCD---it changes the sign of the QCD beta function from positive to negative.}
%
Therefore, the grey bubble in Fig.~\ref{fig:1loopR} is the gluon vacuum polarization at one loop \cite{Grozin:2005yg,Beneke:1998ui,Lautrup:1977hs,Gross:1974jv}
\begin{equation}
\Pi(k^2)=-\frac{\alpha_s(\mu)}{4\pi}\beta_{0}\left(\ln\frac{-k^2}{\mu^2}-\frac{5}{3}\right) \ ,
\end{equation}
where $k$ is the gluon momentum and the 5/3 factor is associated with the $\overline{\text{MS}}$ scheme. Summing the bubble-chains yields a factor modifying the gluon propagator:
\begin{equation}\label{eq:bubble-chain}
\frac{1}{1-\Pi(k^2)}
=\frac{4\pi}{\alpha_s}\int_0^\infty\frac{dw}{\beta_0}e^{-\frac{4\pi w}{\beta_0\alpha_s}+\frac{5}{3}w}\left(\frac{\mu^2}{-k^2}\right)^w \ ,
\end{equation}
which has a form of an inverse Borel transform with the Borel series $\frac{4\pi}{\alpha_s}\left(\frac{e^{\frac{5}{3}}\mu^2}{-k^2}\right)^w$. We will rewrite it as $\left(\frac{\Lambda_{\text{QCD}}^2}{-k^2}\right)^w$ times a factor which is moved to Eq.(\ref{eq:bare_qPDF}). Therefore, to compute the diagrams of Fig.\ref{fig:1loopR}, one just have to replace the gluon propagator by a dressed one to obtain the Borel series:
\begin{figure}
\includegraphics[width=\textwidth]{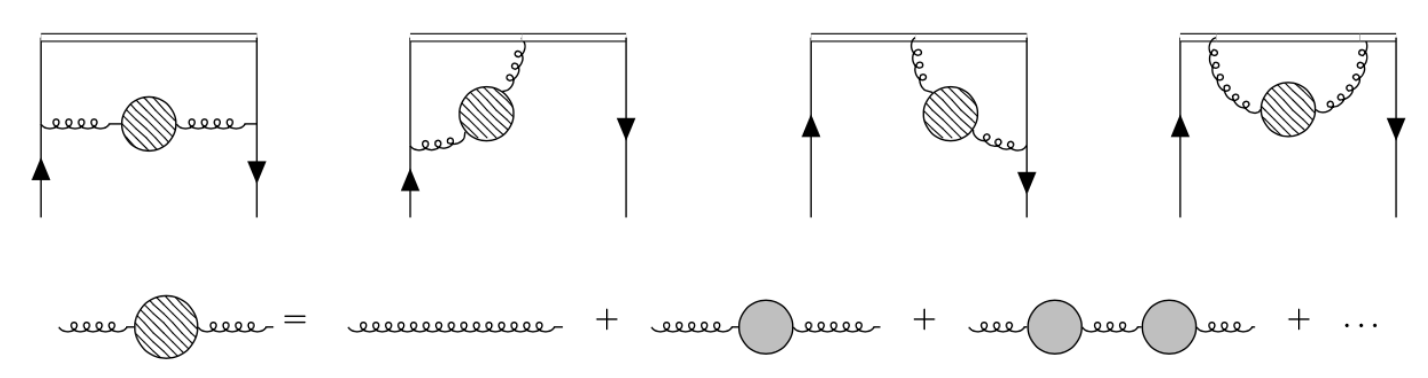}
\caption{Bubble-chain diagrams for PDF and quasi-PDF for single quark states. The single lines are quark propagators, curly lines are gluons propagators and double lines are Wilson lines. The gray bubbles are gluon vacuum polarization diagrams at one loop.}
\label{fig:1loopR}
\end{figure}

\begin{equation}
\label{11}
\frac{1}{-k^2-i\epsilon} \to \frac{(\Lambda_{\text{QCD}}^{2})^w}{(-k^2-i\epsilon)^{1+w}}=\frac{\left(\mu^{2}e^{-\frac{4\pi}{\beta_0\alpha_s}}\right)^w}{(-k^2-i\epsilon)^{1+w}} \ ,
\end{equation}
where Eq.\eqref{eq:running coupling} is used. 

Below we list the unpolarized isovector PDF and quasi-PDF results of these diagrams. The Borel series of the bubble-chain diagrams in Fig.\ref{fig:1loopR} yields the \Blue{{\em quark-level}} quasi-PDF \Blue{$\tilde q(x,\rho)$ (defined similarly to  
Eq.~(\ref{eq:def_quasi-PDF}) but with the external hadronic state replaced by a single quark state of momentum $p_{\mu}$):} 
\begin{equation}\label{eq:bare_qPDF}
B[\tilde{q}(x,\rho)]=8\zeta C_F\left(\frac{\Lambda_{\text{QCD}}^2}{p_z^2}\right)^we^{\frac{4\pi w}{\beta_0\alpha_s}+\frac{5}{3}w}\frac{1}{\sqrt{\pi}}\frac{\Gamma(1/2+w)}{\Gamma(1+w)}[\tilde{f}(x,\rho,w)]_{+} \ ,
\end{equation}
where $\rho=-p^2/p_z^2$ is the off-shellness parameter for external quarks.
$\zeta$ is the quark spinor factor $\zeta=\sum_s\frac{\bar u_s\gamma^zu_s}{4p_z}$. $C_{F}=(N_{c}^{2}-1)/2N_{c}$ with $N_{c}$ the number of colors.
The plus-function  
\begin{equation}
[\tilde{f}(x,\rho,w)]_{+} \equiv \tilde{f}(x,\rho,w)-\delta(x-1) \int_{-\infty}^{\infty} dx' \tilde{f}(x',\rho,w) \ .
\end{equation}
We have computed $\tilde{f}$ 
up to an integral of the Feynman parameter $y$,
\begin{equation}
\label{13}
\begin{aligned}
&\tilde{f}(x,\rho,w)=\\[5pt]
&\int_0^1dy\bigg\{\left(\frac{1+x^2}{1-x}-\frac{\rho}{2(1-x)}\right)\frac{y^w}{[(x-y)^2+y(1-y)\rho]^{w+1/2}}+\frac{wx(1-y)y^{w-1}}{[(x-y)^2+y(1-y)\rho]^{w+1/2}}\\[5pt]
&-\left(\frac{1}{2w-1}\right)\frac{1-x}{|1-x|^{1+2w}}+\frac{1}{2w-1}\frac{1}{1-x}\frac{wy^{w-1}}{[(x-y)^2+y(1-y)\rho]^{w-1/2}}\\[5pt]
&+\frac{\frac{1}{2}+w}{2(1+w)} \frac{\rho(1- x) y^{w+1}}{[(x-y)^2+y(1-y)\rho]^{w+3/2}}-\frac{\left(\frac{3}{2}+w\right)\left(\frac{1}{2}+w\right)}{2(1+w)}\frac{x\rho^2(1-y)y^{w+1}}{[(x-y)^2+y(1-y)\rho]^{w+5/2}}\\[5pt]
&-\frac{\frac{1}{2}+w}{2} \frac{\rho x (1-y)y^{w}}{[(x-y)^2+y(1-y)\rho]^{w+3/2}}+\left(\frac{1}{2w-1}\right)\frac{1}{|1-x|^{1+2w}}\bigg\} \ .
\end{aligned}
\end{equation}

It is easy to see that for positive and finite $\rho$, if $x$ is different from $0$ or $1$ and $w$ is different from $1/2$ or $-1$, then the integrant in the above integral will not diverge.  For $x=0,1$, the $y$ integration diverges when $w$ is bigger than a certain value. To see the poles in $w$, one performs the $y$ integration by assuming $w$ is small enough such that the $y$ integration is finite. Then an analytic continuation of $w$ is performed to look for poles in $w$. After this exercise, no additional pole at $x=0,1$ is found.
The $w=1/2$ pole is corresponding to the UV divergence of the Wilson line self-energy diagram, 
which can be subtracted away by the renormalization procedure, such as the RI/MOM scheme used in this work. Therefore we do not consider it further here. The $w=-1$ pole is not on the path of integration of the inverse Borel transform and hence does not cause any ambiguity.

Setting the external quark on-shell ($\rho \to 0$), Eq.(\ref{13}) for the unphysical region ($x>1$ and $x<0$) becomes
\begin{equation}
\begin{aligned}
\label{eq15}
\tilde{f}(x,0,w)=&\left(\frac{1-x+x^2}{1-x}-wx\right)\frac{1}{|x|^{1+2w}}\frac{1}{1+w}{}_2F_1(1+w,1+2w; 2+w; \frac{1}{x})\\
&+\frac{x}{|x|^{1+2w}}{}_2F_1(w,1+2w; 1+w; \frac{1}{x})+\frac{1}{1-x}\frac{1}{|x|^{1+2w}}\frac{1}{2+w}{}_2F_1(2+w,1+2w; 3+w; \frac{1}{x})\\
&+\frac{1}{2w-1}\frac{1}{|1-x|^{2w+1}}
\ ,
\end{aligned}
\end{equation}
where $_2F_1$ is a hypergeometric function. There is no IR renormalon in this region, except the contribution near $x=1$ which yields the delta function in Eq.(\ref{eq:leading_ren}). The UV renormalon at $w=1/2$ can be subtracted away by the renormalization procedure discussed above and will not be discussed further.

In the physical region $0<x<1$, IR renormalons exist. They correspond to single poles of $\tilde{f}$ at all positive integer values of $w$:
\begin{equation}
\begin{aligned}
\label{eq:f0to1}
\tilde{f}(x,0,w)=&\left(\frac{1-x+x^2}{1-x}-wx\right)\bigg[-\frac{1+(-1)^{2w}}{2x^{w}}\frac{\Gamma(w)\Gamma(1-2w)}{\Gamma(1-w)}\\
&+\left(\frac{-1}{x}\right)^{1+2w}\frac{1}{1+w}{}_2F_1(1+w,1+2w; 2+w; \frac{1}{x})\bigg]\\
&+\frac{1+(-1)^{2w}}{2x^w}\frac{\Gamma(1+w)\Gamma(1-2w)}{\Gamma(1-w)}-\left(\frac{-1}{x}\right)^{2w}{}_2F_1(w,1+2w; 1+w; \frac{1}{x})\\
&+\frac{1}{1-x}\bigg[\frac{1+(-1)^{2w}}{x^{w-1}}\frac{\Gamma(-2w)\Gamma(2+w)}{\Gamma(2-w)}+\left(\frac{-1}{x}\right)^{1+2w}\frac{1}{2+w}{}_2F_1(2+w,1+2w; 3+w; \frac{1}{x})\bigg]\\
&+\frac{1}{2w-1}\frac{1}{|1-x|^{2w+1}} \ .
\end{aligned}
\end{equation}
The leading IR renormalon ambiguity in the inverse Borel transform of $\tilde{q}$ is proportional to the residue of the single pole at $w=1$. 
Eqs.(\ref{eq15}) and (\ref{eq:f0to1}) yield
%
%
\begin{equation}
\begin{aligned}
\label{eq:lim}
(1-x)(1-w)\tilde{f}(x,0,w)&=\theta(x)\theta(1-x)-\frac{(1-w)}{|1-x|^w}+\mathcal{O}((1-w)^2 ) .
\end{aligned}
\end{equation}
When $w\to1$, the term $(1-w)/|1-x|^w$ is zero for $x \neq 1$ but undetermined for $x=1$. It is in fact a delta function since 
%
\begin{equation}
\begin{aligned}
\lim_{w\to1}\int_{1^{-}}^{1^+}dx~\frac{1-w}{|1-x|^w}=1 .
\end{aligned}
\end{equation}
This is consistent with the result of Ref.~\cite{Schwartz:2013pla}: 
\begin{equation}
\label{eq:plus}
\frac{1-w}{|1-x|^w}=\delta(1-x)+\sum_{n=0}^\infty\frac{(1-w)^{(n+1)}}{n!}\left(\frac{\ln^n|1-x|}{|1-x|}\right)_+ .
\end{equation}
Eq.~\eqref{eq:lim} yields
\begin{equation}
\begin{aligned}
\label{eq:leading_ren}
&\lim_{\substack{w\to1}}(1-x)(1-w)\tilde{f}(x,0,w)=\theta(x)\theta(1-x)-\delta(1-x) ,
\end{aligned}
\end{equation}
and the leading IR renormalon ambiguity in the inverse Borel transform of $\tilde{q}$ is thus given by  
\begin{equation}
\begin{aligned}
-\frac{\pi}{\beta_0}e^{-\frac{4\pi}{\beta_0\alpha_s}}\text{Res}[B[\tilde{q}(x),0]]\bigg|_{w=1}&=\frac{\pi}{\beta_0}e^{5/3}\frac{\Lambda_{\text{QCD}}^2}{p_z^2}4C_F\left[\frac{\theta(x)\theta(1-x)-\delta(x-1)}{1-x}\right]_+
\ .
\label{eq:renormalon}
\end{aligned}
\end{equation}
\Blue{We stress that the IR renormalon is associated with the IR divergence generated when all the particles in the loop become on-shell simultaneously. This is purely an IR effect. Therefore, there is no dependence on the UV regulator in Eq.~(\ref{eq:renormalon}).}

\begin{figure}[t]
\includegraphics[width=.7\textwidth]{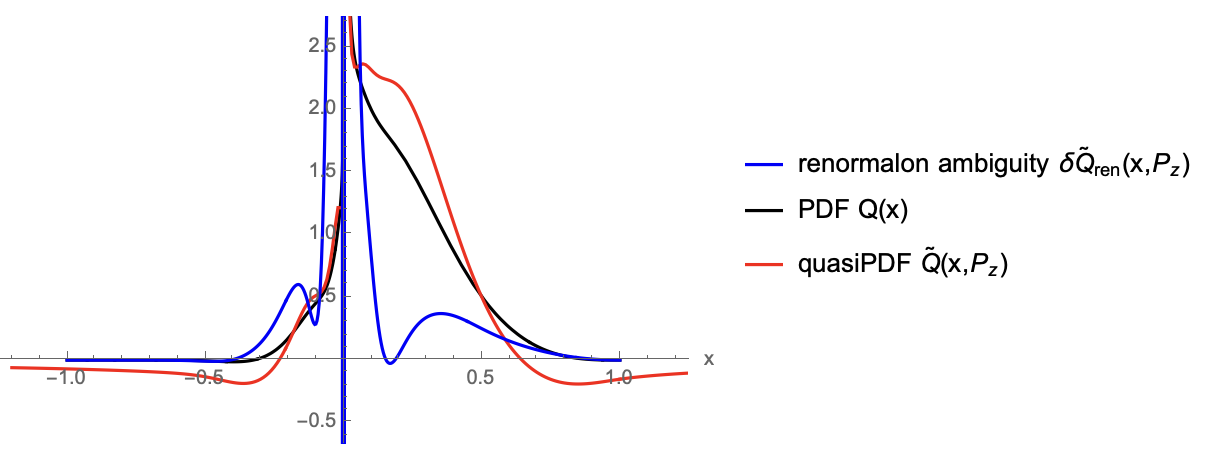}
\hfill
\caption{\Blue{The leading renormalon ambiguity $\delta\tilde{Q}_{ren}(x, P_z)$ in the proton isovector quasi-PDF
together with the corresponding PDF, $Q(x)$, and the one-loop quasi-PDF $\tilde{Q}(x, P_z)$, all in the $\overline{\text{MS}}$ scheme. $\delta\tilde{Q}_{ren}(x, P_z)$ is generated with Eq.~(\ref{eq:Qrenorm}), $P_{z}=1.5$ GeV and $Q(x)$ from Ref. \cite{Owens:2012bv} (CJ12). 
$\tilde{Q}$, which is also shown in Fig.~\ref{fig:qPDF}, is generated by convoluting $Q(x)$ with the one-loop matching kernel.
$\delta\tilde{Q}_{ren}(x, P_z)$ has a large negative contribution near $x=0$ such that the area under the curve is zero. This is a consequence of quark number conservation.
The size of $\delta\tilde{Q}_{ren}(x, P_z)$ is bigger than $\tilde{Q}(x, P_z)$ for almost the entire region of $-0.3 \lesssim x \lesssim 0.1$. 
}}
\label{fig:Qren}
\end{figure}

\Blue{After discussing the quasi-PDF, we now turn to the PDF. The quark-level PDF $q(x,\frac{\mu^2}{-p^2})$ is defined similarly to Eq.~(\ref{eq:def_PDF}) but with the external hadronic state replaced by a single  quark state of momentum $p_{\mu}$.} The Borel series of the bubble-chain diagrams in Fig.\ref{fig:1loopR} for quark-level PDF can be obtained by taking $p_z$
to infinity before integrating the spacial loop momenta in other directions. The result is 
\begin{equation}\label{14}
B[q(x,\frac{\mu^2}{-p^2})]=8\zeta C_F\left(\frac{\Lambda_{\text{QCD}}^2}{-x(1-x)p^2}\right)^we^{\frac{4\pi w}{\beta_0\alpha_s}+\frac{5}{3}w}\frac{\Gamma(w+\epsilon)}{\Gamma(1+w)}[f(x,\epsilon,\frac{\mu^2}{-p^2}, w)]_{+} \ ,
\end{equation}
where 
\begin{equation}
\begin{aligned}
f(x,\epsilon,\frac{\mu^2}{-p^2},w)=&\left(\frac{4\pi\mu^2}{-x(1-x)p^2}\right)^\epsilon x^w\bigg\{(1-\epsilon)\left[1-x+w(1-x)-(w+\epsilon)\right]\\[5pt]
&+\frac{2x}{1-x}-\left[\frac{(w+1+\epsilon)(w+\epsilon)}{2(w+1)}\frac{1}{1-x}-\frac{w+\epsilon}{2}-\frac{w+\epsilon}{2(w+1)}\right]\bigg\} \ ,
\end{aligned}
\end{equation}
and where the dimensional regularization parameter $\epsilon$ is kept because we want to perform a $w$ expansion later and this helps to keep the factor $\Gamma(w+\epsilon)$ regulated. 
%
%
The quark-level PDF does not have any IR renormalon ambiguity as there is no pole in the Borel series in the positive $w$ real axis. This result is expected, since the PDF is corresponding to quasi-PDF in the $p_{z} \to \infty$ limit and the leading IR renormalon, which is is proportional to $1/p_{z}^{2}$ as shown in Eq.~(\ref{eq:renormalon}), vanishes in this limit.


We can now study how this leading renormalon ambiguity \Blue{$\delta\tilde{Q}_{ren}$} 
affects the quasi-PDF of a hadron, $\tilde{Q}$, through the matching kernel of Eq.(\ref{eq:matchingX}). The kernel can be converted from Eq.\eqref{eq:renormalon} by replacing the quark momentum $p_z$ with $yP_z$, $P_z$ being the hadron momentum, and replacing $x$ with $x/y$, such that
\begin{equation}
\begin{aligned}
\delta\tilde{Q}_{ren}(x, P_z)=&\frac{\pi}{\beta_0}e^{5/3} C_F\frac{\Lambda_{\text{QCD}}^2}{P_z^2}\int_{-1}^1\frac{dy}{|y|y^{2}}\left[\frac{\theta(1-\frac{x}{y})\theta(\frac{x}{y})-\delta(1-\frac{x}{y})}{1-\frac{x}{y}}\right]_{+}Q(y)\\ 
=&\frac{\pi}{\beta_0}e^{5/3} C_F\frac{\Lambda_{\text{QCD}}^2}{x^2P_z^2}\int_{|x|}^1 d\xi \left[\frac{ 1}{1-\xi}-\delta(1-\xi)\frac{ 1}{1-\xi}\right]_+ \xi Q(x/\xi) \\
=&\frac{\pi}{\beta_0}e^{5/3} C_F\frac{\Lambda_{\text{QCD}}^2}{x^2P_z^2}\int_{0}^1 d\xi \left[\frac{ 1}{1-\xi}-\delta(1-\xi)\frac{ 1}{1-\xi}\right] \left(\xi Q(x/\xi)-Q(x)\right) \\
=&\frac{\pi}{\beta_0}e^{5/3} C_F\frac{\Lambda_{\text{QCD}}^2}{x^{2}P_z^2}\left\{\int_{0}^1 d\xi  \frac{1}{1-\xi}\left[\xi Q(x/\xi)-Q(x)\right]+Q(x)-xQ'(x)\right\} \ ,
\end{aligned}
\label{eq:Qrenorm}
\end{equation}
where the integration limit $|x|$ in the second line can be replaced by $0$ since $Q(x)$ vanishes outside of $x=[-1,1]$ and 
we have expanded $(\xi Q(x/\xi)-Q(x))/(1-\xi)$ around $\xi =1$ before acting it with $\delta(1-\xi)$ in the fourth line.

The plus-function in Eq.~(\ref{eq:Qrenorm}) ensures quark number conservation, such that $\int dx \ \delta\tilde{Q}_{ren}= 0$. One can show the first two terms in the last line of Eq.~(\ref{eq:Qrenorm}) cancel after integrating over $x$, and the last two terms are proportional to 
$\frac{d}{dx}\left(Q(x)/x\right)$, which yields vanishing boundary terms after integrating over $x$. Our Eq.~(\ref{eq:Qrenorm}) is the same as the coordinate space calculation in Ref.~\cite{Braun:2018brg}.\footnote{Except our prefactor of $Q'(x)$ is $-x$ rather than  $-|x|$ as in Ref.~\cite{Braun:2018brg}. But if one starts from the first line of Eq. (61) of Ref.~\cite{Braun:2018brg}, then the prefactor $Q'(x)$ will be $-x$. So this should be a typo of Ref.~\cite{Braun:2018brg}.}

\Blue{In Fig.\ref{fig:Qren}, the leading renormalon ambiguity $\delta\tilde{Q}_{ren}(x, P_z)$ in the proton isovector quasi-PDF, is shown,  
together with the corresponding PDF, $Q(x)$, and the one-loop quasi-PDF $\tilde{Q}(x, P_z)$, all in the $\overline{\text{MS}}$ scheme. 
Numerically, $\delta\tilde{Q}_{ren}(x, P_z)$ is a significant effect. Its size is bigger than $\tilde{Q}(x, P_z)$ for almost the entire region of $-0.3 \lesssim x \lesssim 0.1$.
%
In the computation, we have used $Q$ from the proton isovector (i.e. the $(u-d)$ quark combination) PDF extracted by the CTEQ-JLab collaboration (CJ12) \cite{Owens:2012bv}. $\tilde{Q}$, which is also shown in Fig.~\ref{fig:qPDF}, is computed with $Q(x)$ convoluted with the one-loop matching kernel. The renormalon ambiguity $\delta\tilde{Q}_{ren}$ of $\tilde{Q}$ is computed with Eq.~(\ref{eq:Qrenorm}) with $P_{z}=1.5$ GeV and $\Lambda_{\text{QCD}} =0.254 $ GeV. The areas under $Q$ and $\tilde{Q}$ are both one, the isovector charge, while the area under $\delta\tilde{Q}_{ren}$ is zero, with a large negative contribution near $x=0$. This is the result of fermion number conservation. 
}

\Blue{The singular behavior near $x=0$ requires the introduction of an IR regulator
\begin{equation}
Q(x) \to Q(x) \theta(|x|-|\bar \epsilon|) .
\end{equation}
When $\bar \epsilon \to 0$ is taken at the end of the calculation, 
the contribution for $|x| <|\bar \epsilon|$ becomes $\delta(x)$ with a divergent prefactor. This illusive contribution could be easily overlooked in a numerical analysis. But it is critical for the demonstration of quark number conservation, which is a property of the bubble-chain diagrams that we considered.}

\Blue{Both $\delta\tilde{Q}_{ren}(x, P_z)$ and $Q(x)$ have support in $[-1,1]$. 
When $x \to 0$ or $x \to 1$, 
\begin{equation}\label{XX}
\delta\tilde{Q}_{ren}(x, P_z)  \propto \frac{\Lambda_{\text{QCD}}^2}{x^{2 }(1-x)P_z^2} Q(x) ,
\end{equation}
which is consistent with Ref. \cite{Braun:2018brg}. The quasi-PDF $\tilde{Q}(x, P_z)$, however, has support for all values of $x$. Hence $\delta\tilde{Q}_{ren}$ has no $1/(1-x)$ enhancement compared to $\tilde{Q}$. Therefore,
\begin{equation}\label{XXX}
\delta\tilde{Q}_{ren}(x, P_z)  \propto  \frac{\Lambda_{\text{QCD}}^2}{x^{2 }P_z^2} \tilde{Q}(x, P_z) .
\end{equation}
  }

\Blue{Since the renormalon ambiguity 
discussed above should be cancelled by power corrections, one can rewrite Eq.(\ref{eq:matchingX}) as 
\begin{equation} \label{eq:matchingXX}
\tilde{Q}(x, P_z, \mu')=\int_{-1}^1 \frac{dy}{|y|} Z(\frac{x}{y},y P_z,\mu',\mu) Q(y, \mu) +\mathcal{O}(\frac{1}{x^{2}P_z^2},\frac{\delta(x)}{P_z^2}) \ ,
\end{equation}
where the first power correction has no $1/(1-x)$ factor and the second power correction has a divergent prefactor. 
}

\section{Removing the leading renormalon ambiguity}

\Blue{Although the bubble-chain diagrams employed in the analysis above might not be numerically the dominant contribution in QCD, we still expect that they will generate all structures of the renormalon ambiguity \cite{Brown:1991ic}. Hence Eq.(\ref{eq:matchingXX}) should be quite robust and is expected to be free from the shortcoming of the bubble-chain diagram analysis. }
  
\Blue{The R-scheme proposed in Ref.~\cite{Hoang:2009yr} is designed to remove the leading renormalon ambiguity in a general operator product expansion. In our case, the leading renormalon ambiguity  in the matching factor $Z$ and the leading power corrections are both proportional to $1/P_z^2$ in Eq.(\ref{eq:matchingXX}). These $1/P_z^2$ terms can be removed in the following combination: 
\begin{equation}
\begin{aligned}
\label{28}
\tilde{Q}_{R}(x, P_z,P^\prime_z,\Lambda')=& \frac{P_z^2 \tilde{Q}(x, P_z,\Lambda')-P_z^{\prime 2} \tilde{Q}(x, P^\prime_z,\Lambda')}{P_z^2-P_z^{\prime 2}} \\
=&\int_{-1}^{1}\frac{dy}{|y|}\left[\frac{P_z^2 Z(\frac{x}{y}, y P_z,\Lambda',\mu)-P_z^{\prime 2}Z(\frac{x}{y}, y P^\prime_z,\Lambda',\mu)}{P_z^2-P_z^{\prime 2}}\right]Q(y, \mu)+\mathcal{O}(\frac{\alpha_{s}\ln{(P_{z}/P^{\prime}_{z})}}{P_z^2-P_z^{\prime 2}},\frac{1}{P_z^2 P_z^{\prime 2}}) \ .
\end{aligned}
\end{equation}
%
The remaining power corrections can in principle be removed in a similar way as well. The logarithmic term comes from the $(\alpha_{s}\ln{P_{z}})/P_z^2$ correction in the kernel $Z$ which can be recasted to anomalous dimension $\gamma$ such that the leading renormalon ambiguity in $Z$ scales as $1/P_z^{(2+\gamma)}$. This power can be removed by replacing $2 \to 2+\gamma$ in the square brackets. Now the remaining power correction is $\mathcal{O}(1/P_z^{2} P_z^{\prime 2})\sim\mathcal{O}(1/P_z^{4})$. So the procedure can be applied again to remove this power correction. However, the computation of the anomalous dimensions beyond the bubble-chain diagrams could be a challenge in removing subleading renormalon ambiguities with this method.  }

\Blue{Ref.~\cite{Braun:2018brg} also proposed a method to remove the leading renormalon ambiguity by replacing the $\gamma^{z}$ of
Eq.(\ref{eq:def_quasi-PDF}) with a combination of $\gamma^{z}$ and $\gamma^{t}$. The idea is that PDF can be extracted using either $\gamma^{z}$ or $\gamma^{t}$ matrix element. However, the leading renormalon ambiguity contributes differently in the two cases. By choosing a specific combination of the $\gamma^{z}$ and $\gamma^{t}$ matrix elements, the leading renormalon ambiguity is largely canceled. However, knowing what combination to use is difficult beyond the bubble-chain approximation. Also, this proposal has another technical issue. On a lattice, the matrix element using $\gamma^{z}$ will mix with another operator of the same dimension that needs to be removed, while the matrix element using $\gamma^{t}$ does not have this problem. Therefore, typically $\gamma^{t}$ is used in lattice computations. However, for the purpose of this work, both of the choices are equivalent since the analysis is performed in the continuum.}


\section{Bubble-Chain Contributions in Fixed Order Perturbations} 
\label{pert}

\Blue{In the previous section, we have seen that the leading IR renormalon ambiguity of Eq.(\ref{eq:Qrenorm}) 
yields $1/P_z^{2}$ power corrections of Eq.(\ref{eq:matchingXX})
with singular prefactors after summing the bubble-chain diagrams to all orders in $\alpha_{s}$. These $1/P_z^{2}$ terms are canceled in the R-scheme by design in Eq.(\ref{28}). It is natural to ask, if the cancellation works when the series is summed to all orders in $\alpha_{s}$, can the cancellation also happen if the series is truncated to a fixed order? 
It seems the answer should be yes if the cancelation works for the infinite series and for different values of $\alpha_{s}$ (within the radius of convergence), then the cancelation ought to happen at each order in  $\alpha_{s}$ as well.}

\begin{figure}[t]
\centering
\begin{subfigure}{.47\textwidth}
\centering
\includegraphics[width=\textwidth]{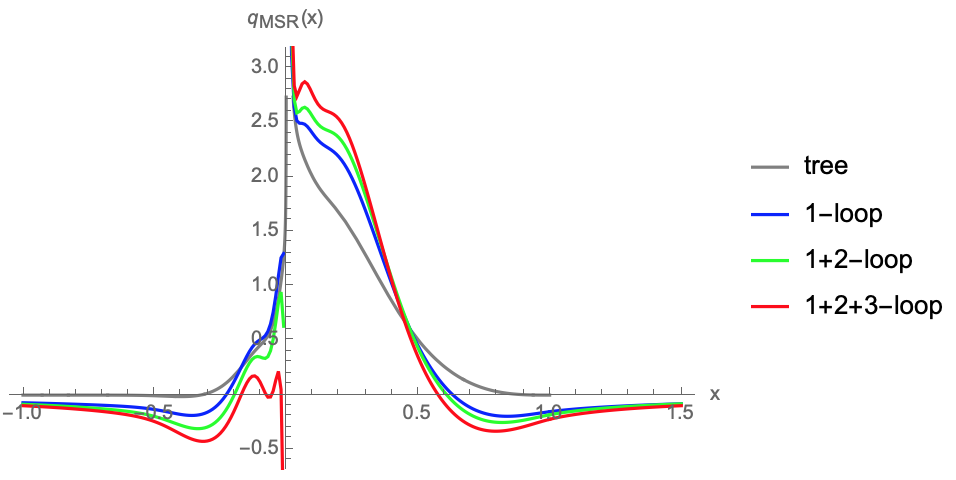}
\caption{}
\end{subfigure}
\hfill
\begin{subfigure}{.47\textwidth}
\centering
\includegraphics[width=\textwidth]{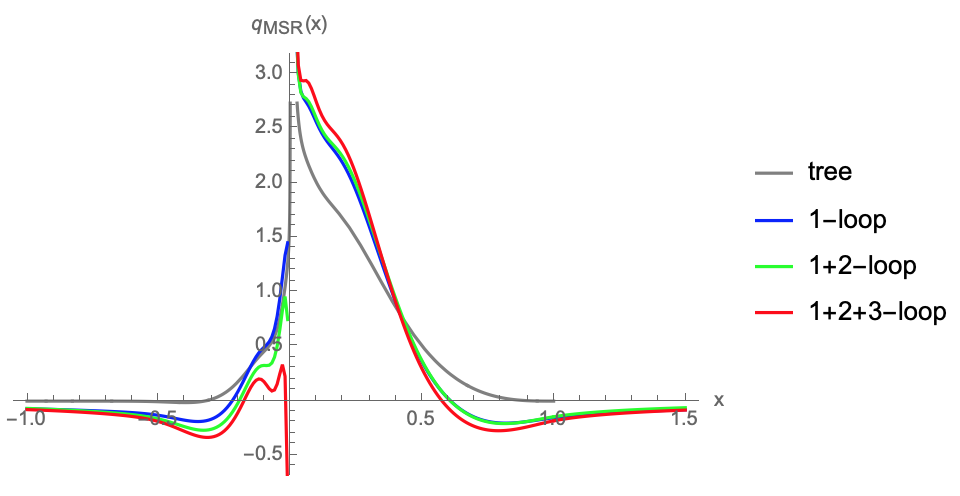}
\caption{}
\end{subfigure}
\caption{ Isovector proton quasi-PDFs in $\overline{\text{MS}}$ with R-scheme for (a) $P^{\prime}_z=1$ GeV and (b) $P^{\prime}_z=3$ GeV,  $\alpha_s=0.283$, $P_z=1.5$ GeV, and $\mu=3$ GeV. Rapid convergence of the series expansion is seen in (b).}
\label{fig:RPlot}
\end{figure}

\Blue{To answer this question, we need multi-loop corrections to the PDF and quasi-PDF. However, complete results beyond one-loop does not exist yet because of their technical difficulty. Therefore, we resort to large $n_{F}$ expansion again and only include the bubble-chain diagrams.  As we will see in Fig.~\ref{fig:RPlot}(b), the R-scheme could indeed  introduce large cancellation of the bubble-chain diagrams at each order in  $\alpha_{s}$. This is consistent with what Ref.~\cite{Hoang:2009yr} advocated and demonstrated that R-scheme indeed improved the convergences of QCD perturbation.
}

Now we show how to extract the fixed order result from the bubble summed result. The  bubble-chain contribution to the $(n+1)$-th loop quasi-PDF in the $\overline{\text{MS}}$ scheme can be extracted by expanding the Borel series of Eq.(\ref{eq:bare_qPDF}) 
in powers of $w$: 
\begin{equation}
\label{29}
\begin{aligned}
\tilde{q}^{(n+1)}(x,\rho)=&\left(\frac{\alpha_s}{4\pi}\right)^{n+1}\beta_0^{n}\left(\frac{d}{dw}\right)^{n}B[\tilde{q}](w)\bigg|_{w=0}\\[5pt]
=&8\zeta C_F\left(\frac{\alpha_s}{4\pi}\right)^{n+1}\beta_0^{n}\sum_{k=0}^n\binom{n}{k}\left(\ln\frac{\mu^2e^{5/3}}{4x^2p_z^2}\right)^{n-k}\tilde{C}_{k+1}(x, \rho)  \ ,
\end{aligned}
\end{equation}
where 
\begin{equation}
\tilde{C}_{k+1}(x, \rho)=\frac{d^{k}}{dw^{k}}\bigg\{\frac{(4x^2)^w\Gamma(1/2+w)}{\sqrt{\pi}\Gamma(1+w)}\left[\tilde{f}(x,\rho,w)\right]_{+}\bigg\}\bigg|_{w=0} \ .
\end{equation}

The bubble-chain contribution to the $(n+1)$-th loop PDF in the $\overline{\text{MS}}$ scheme is
\begin{equation}
\begin{aligned}
q^{(n+1)}(x,\frac{\mu^2}{-p^2})=&\left(\frac{\alpha_s}{4\pi}\right)^{n+1}\beta_0^{n}\left(\frac{d}{dw}\right)^{n}B[q](w)\bigg|_{w=0}\\[5pt]
=&8\zeta C_F\left(\frac{\alpha_s}{4\pi}\right)^{n+1}\beta_0^{n}\sum_{k=0}^n\binom{n}{k}\left(\ln\frac{\mu^2e^{5/3}}{-(1-x)p^2}\right)^{n-k}C_{k+1}(x, \frac{\mu^2}{-p^2}) \ ,
\end{aligned}
\end{equation}
where 
\begin{equation}
C_{k+1}(x, \frac{\mu^2}{-p^2})=\frac{d^{k}}{dw^{k}}\bigg\{\frac{\Gamma(w+\epsilon)}{x^w\Gamma(1+w)} \left[ f(x,\epsilon, \frac{\mu^2}{-p^2},w)\right]_{+}
\bigg\}\bigg|_{w=0, \epsilon \to 0} \ . 
\end{equation}

The results for $\tilde{q}^{(n)}$, $q^{(n)}$, together with the $\overline{\text{MS}}$ to $\overline{\text{MS}}$
matching kernel 
\begin{equation}
\label{Zn}
\begin{aligned}
Z^{(n)}(\xi, \frac{p_z^2}{\mu^2})=&\frac{1}{4\zeta}\left[\tilde{q}^{(n)}(\xi,\rho \to 0)-q^{(n)}(\xi,-\mu^2/p^2)\right]\\[5pt]
\end{aligned}
\end{equation}
 for $n=1,2,3$ are listed in the Appendix.

 These matching kernels are free from IR divergence from diagram-by-diagram cancellations. Note that we should have also included the convolution term $Z^{(1)}\otimes q^{(1)}$ in the $Z^{(2)}$ kernel and the $Z^{(2)}\otimes q^{(1)}+Z^{(1)}\otimes q^{(2)}$ term in the $Z^{(3)}$ kernel were we to include all the $n$-loop diagrams rather than just the bubble-chain diagrams. But here we only investigate how the bubble-chain diagrams which could give rise to the renormalon ambiguities converge at higher loops. 
 
 \begin{figure}[t]
\centering
\begin{subfigure}{.45\textwidth}
\centering
\includegraphics[width=\textwidth]{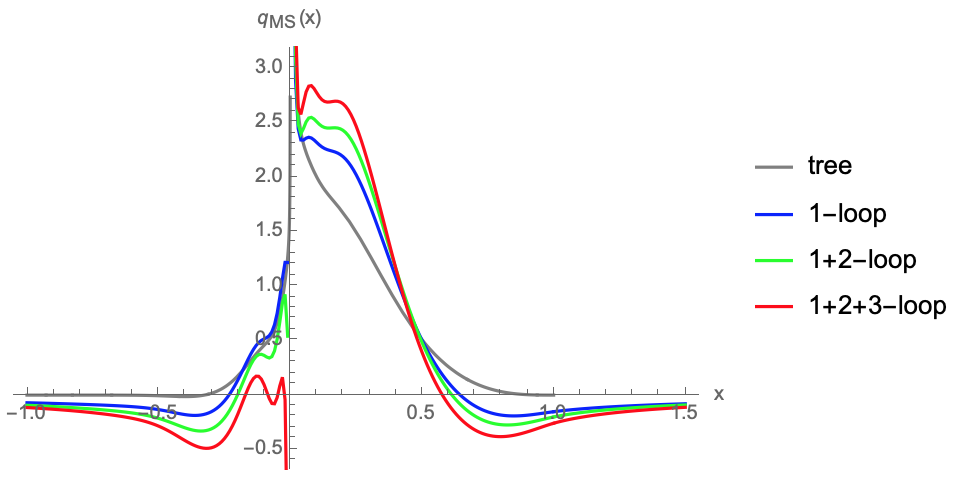}
\end{subfigure}
\hfill
\caption{Isovector proton quasi-PDF's in the $\overline{\text{MS}}$ scheme derived from convoluting the CJ12 proton PDF~\cite{Owens:2012bv} (shown as the tree level result) with the matching kernels computed with bubble-chain diagrams with $\alpha_s=0.283$, $P_z=1.5$ GeV, $\mu=3$ GeV. The convergence of the series expansion is slow.}
\label{fig:qPDF}
\end{figure}


\Blue{To check the effect of the bubble-chain contribution at the $n$-loop, we convolute the CJ12 proton isovector PDF~\cite{Owens:2012bv} with $Z^{(n)}$ 
of Eq.(\ref{Zn}) to obtain the corresponding isovector quasi-PDF of the proton at $n$-loop in the $\overline{\text{MS}}$ scheme.  The result up to 3-loops is shown in Fig.\ref{fig:qPDF}. We see that the two-loop contribution is smaller than one-loop but about the same size as the three-loop. Therefore the convergence is already slow at three-loop for the $\overline{\text{MS}}$ quasi-PDF. However, as shown in Fig.~\ref{fig:RPlot}, the convergence is much better when we use the R-scheme of Eq.(\ref{28}), especially when a larger $P_z^{\prime}$ is used. The need to use a large $P_z^{\prime}$ can be understood because for a fixed order expansion, the kernel only has powers of $\ln P_z^{}\prime$ dependence but not $1/P_z^{\prime 2}$ dependence. Hence when we take $P_z^{\prime} \to 0$, R-scheme reduces to the usual  $\overline{\text{MS}}$ scheme shown in Fig.\ref{fig:qPDF} and has slow convergence. Therefore, larger $P_z^{\prime}$ tends to yield larger cancellation of the bubble-chain diagrams.}
 
After establishing the power of R-scheme in removing the bubble-chain diagram contributions in 2- and 3-loop diagrams, we turn to the RI/MOM
scheme which is typically used in lattice QCD and see how bubble-chain diagrams contribute in this scheme. In RI/MOM, all the loop corrections to 
 the matrix element of a single quark state with momentum $p_{\mu}$ 
 are subtracted non-perturbatively at an off-shell kinematics $-p^2=\mu^2_R$ and $p_z=p^z_R$. In momentum space, this amounts to the subtraction  \cite{Stewart:2017tvs}
\begin{equation}
\label{qR}
\tilde{q}^{(n)}_R(x,\rho,\mu_R^2/(p^z_R)^2)=\tilde{q}^{(n)}(x,\rho)-|\eta|\tilde{q}^{(n)}(1+\eta(x-1),\mu_R^2/(p^z_R)^2) \ ,
\end{equation}
with $\eta=p_z/p^z_R$. The RI/MOM renormalized quasi-PDF $\tilde{q}_{R}^{(n)}$ is UV finite, so it does not matter what UV regulator is used to compute $\tilde{q}^{(n)}$ of Eq.(\ref{qR}). The UV regulator will be removed at the end to obtain $\tilde{q}_{R}^{(n)}$.
We will just replace $\tilde{q}^{(n)}$ in Eq.(\ref{Zn}) by $\tilde{q}_{R}^{(n)}$ to get the $\overline{\text{MS}}$ to RI/MOM matching kernel. \Blue{It is worth commenting that all the IR renormalon ambiguity comes from the first term in Eq.(\ref{qR}). The second term, the counterterm, takes the external quark off-shell, hence the Feynman diagrams do not experience IR divergence using the gluon propagator of Eq.~(\ref{11}). Therefore, the subtraction in the RI/MOM scheme does not add additional renormalon ambiguity.}

\begin{figure}[t]
\centering
\begin{subfigure}{.45\textwidth}
\centering
\includegraphics[width=\textwidth]{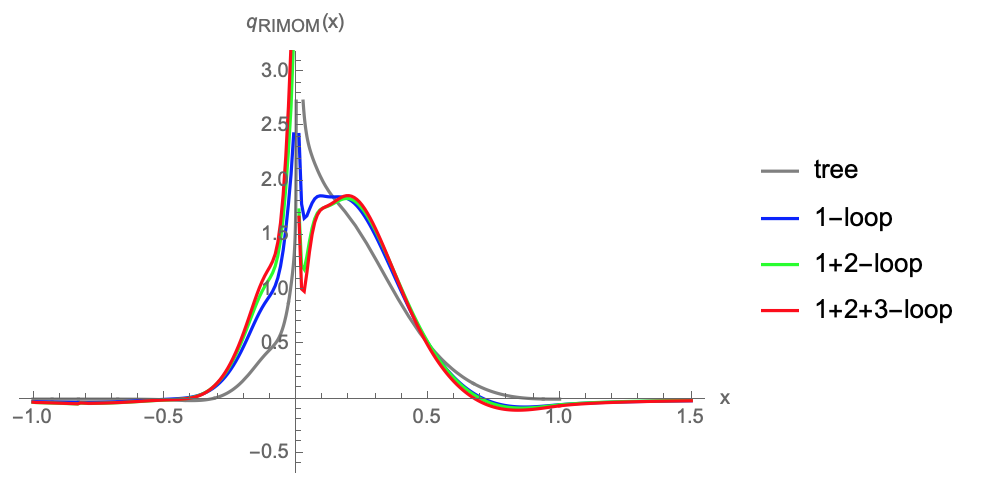}
\end{subfigure}
\caption{Isovector proton quasi-PDF's derived from convoluting the proton PDF of Ref.\cite{Owens:2012bv} with the matching kernels from bubble chain diagrams
for (a)  quasi-PDF's in the $\overline{\text{MS}}$ scheme (b) RI/MOM renormalized quasi-PDFs, with $\alpha_s=0.283$, $P_z=1.5$ GeV, $\mu=3$ GeV, $\mu_R=2.4$ GeV, and $p^z_R=1.201$ GeV.}
\label{fig:qPDFplotRI}
\end{figure}

\Blue{The effect of applying the RI/MOM renormalization to the proton isovector quasi-PDF up to three-loop bubble-chain diagrams is shown in Fig.\ref{fig:qPDFplotRI}. 
Convergence is seen for the whole range of $x$---the possible slow convergence due to the renormalon effect does not appear up to three-loops in the 
RI/MOM scheme. Of course, this result is not conclusive for QCD due to the non-convergence problem of the $1/n_{f}$ expansion mentioned above. However, the convergence pattern in the large $n_{f}$ world is still interesting on its own. It gives some hope that QCD might have a similar convergence pattern. But one can only know after carrying out the complete higher loop calculations.}

\Blue{Finally, we comment on whether we can take advantage of choosing the BLM scales \cite{Brodsky:1982gc} to improve the convergence of the series expansion. The BLM approach is based on the observation that  
\begin{equation}
\label{BLM}
\alpha_{\overline{\text{MS}}}(\mu e^{-2 D})= \alpha_{\overline{\text{MS}}}(\mu)\left(1+ D \beta_{0} \frac{\alpha_{\overline{\text{MS}}}}{\pi}+\cdots\right) .
\end{equation}
Hence by choosing a different renormalization scale for each order in the expansion, one can completely remove all the $ \alpha_{\overline{\text{MS}}}^{n+1}\beta_{0}^{n}$ dependence in the expansion to speed up the convergence. 
Interestingly, 
the $\alpha_{\overline{\text{MS}}}^{n+1}\beta_{0}^{n}$ dependence is exactly what the bubble-chain diagrams yield (see e.g. Eq.(\ref{29})). Hence the BLM approach can cancel the renormalon ambiguity from the bubble-chain diagrams to improve the convergence. However, unlike the factor $D$ of Eq.(\ref{BLM}) is a constant, 
the prefactors  of our matching kernel depend on the momentum fraction. It is not clear how to absorb them into renormalization scales. Even if we proceed by demanding the bubble-chain diagrams be canceled only for a specific momentum fraction, we still need to benchmark the convergence of this approach with the complete multi-loop results which are not yet available. Therefore, we do not pursuit this program here in this work.}

\section{Conclusion} 

We have investigated the renormalon ambiguity in the flavor non-singlet quasi-PDF of a hadron. 
\Blue{We follow the usual practice to study the diagrams in the large $n_{f}$ (the number of fermion flavors) limit with powers of $(\alpha_{s } n_{f})$ summed to all orders \cite{Beneke:1998ui}. Although QCD is not asymptotically free in this limit, the qualitative features of the renormalon ambiguity are expected to remain in QCD \cite{Brown:1991ic}. Also, taking the $\beta$ function of QCD as a reference, it might be possible that the higher order $1/n_{f}$ corrections of our calculation are about the same magnitude as the leading order. In that case, our large $n_{f}$ analysis could still be useful although conclusive results can only be drawn with explicit full QCD calculations.}

The 
bubble-chain diagrams were computed in momentum space and the result agreed with the coordinate space computation of Ref. \cite{Braun:2018brg} up to one possible typo in a prefactor. We confirmed the assertion of Ref. \cite{Braun:2018brg} that the leading IR renormalon ambiguity from bubble-chain diagrams was an $\mathcal{O}(\Lambda^2_{\text{QCD}}/x^{2}P_z^2)$ effect on the quasi-PDF. \Blue{In addition, we have also found an $\mathcal{O}(\delta(x)\Lambda^2_{\text{QCD}}/P_z^2)$ term with a divergent prefactor such that quark number conservation is not broken.}  
This ambiguity is supposed to be cancelled by power corrections associated with higher-twist contributions. Hence, if the power corrections were not included in the analysis, \Blue{then the error was $\mathcal{O}(\Lambda^2_{\text{QCD}}/x^{2}P_z^2,\delta(x)\Lambda^2_{\text{QCD}}/P_z^2)$, which is quite significant at small enough $x$, rather than $\mathcal{O}(\Lambda^2_{\text{QCD}}/P_z^2)$ from dimensional analysis.} 

\Blue{To remove this leading IR renormalon ambiguity, we have investigated the proposed R-scheme~\cite{Hoang:2009yr}, which is designed to cancel all the $\mathcal{O}(1/P_z^2)$ contributions. This cancelation does not rely on the bubble-chain approximation. It is applicable in QCD. Furthermore, since the R-scheme cancel the renormalon ambiguity non-perturbatively for different values of $\alpha_{s}$, it is possible that the cancellation also happens at each order in $\alpha_{s}$ to improve the convergence of QCD perturbation~\cite{Hoang:2009yr}. Because the full multi-loop results in QCD are not available, we use the bubble-chain diagrams to demonstrate this up to three-loops in Fig.~\ref{fig:RPlot}(b).}
 
\Blue{After establishing the power of R-scheme in removing the bubble-chain diagram contributions, we turn to the RI/MOM scheme which is typically used in lattice QCD and see how bubble-chain diagrams contribute to the proton isovector quasi-PDF in Fig.\ref{fig:qPDFplotRI}. 
Convergence is seen for the whole range of $x$---the possible slow convergence due to the renormalon effect does not appear up to three-loops in the 
RI/MOM scheme. Of course, this result is not conclusive for QCD due to the non-convergence problem of the $1/n_{f}$ expansion mentioned above. But it gives some hope that QCD might have a similar convergence pattern. The final result can only be known after carrying out the complete higher loop calculations.}


\section*{Acknowledgments}
We thank Iain Stewart for early involvement of this project and Jian-Hui Zhang for useful discussions. This work is partly supported by the Ministry of Science and Technology, Taiwan, under Grant No. 108-2112-M-002-003-MY3 and the Kenda Foundation.

\appendix

\section{Loop Expansion of the bubble-chain Diagrams}

In this Appendix, we show the $n$-th loop correction to the lightcone PDF $q^{(n)}$ and quasi-PDF $\tilde{q}^{(n)}$ for a single quark state. We only show the result in the $\overline{\text{MS}}$ scheme. The non-perturbative renormalization of the quasi-PDF in the RI/MOM scheme can be constructed using Eq.(\ref{qR}).

\subsection{One-loop result}
The 1-loop correction to quasi-PDF of a single quark state is
\begin{equation}
\label{A1}
\begin{aligned}
\tilde{q}^{(1)}(x,\rho)=8\zeta C_F\frac{\alpha_s}{4\pi}\tilde{C}_1(x,\rho) \ ,
\end{aligned}
\end{equation}
where 
\begin{equation}
\begin{aligned}
\tilde{C}_1(x,\rho)=&\frac{1}{\sqrt{1-\rho}}\left(\frac{1+x^2}{1-x}-\frac{\rho}{2(1-x)}\right)\ln\frac{|1-x|+|x|+\sqrt{1-\rho}}{|1-x|+|x|-\sqrt{1-\rho}}+\frac{x}{|x|}+\frac{|1-x|-|x|}{1-x}-\frac{1}{|1-x|}\\[5pt]
&-x\frac{(\rho-2+2x)|x|-(\rho-2x)|1-x|}{(\rho-4x+4x^2)|x(1-x)|}+x\frac{(\rho-2x+2x^2)-2|x(1-x)|}{(\rho-4x+4x^2)|1-x|}\\[5pt]
&-\frac{x}{2}\left[\frac{\rho^2-4\rho x+8x^2-8x^3}{(\rho-4x+4x^2)^2|x|}+\frac{\rho^2-4\rho+4\rho x+8x(1-x)^2}{(\rho-4x+4x^2)^2|1-x|}\right]\\[5pt]
&+\frac{x}{2}\frac{(\rho-2+2x)|x|+(\rho-2x)|1-x|}{(\rho-4x+4x^2)|x(1-x)|}+\frac{(\rho-2x+2x^2)-2|x(1-x)|}{2(\rho-4x+4x^2)|1-x|}\\[5pt]
&-x\frac{(\rho-2x+2x^2)-2|x(1-x)|}{(\rho-4x+4x^2)|1-x|}\\[5pt]
=&\begin{cases}
\frac{1}{\sqrt{1-\rho}}\left(\frac{1+x^2}{1-x}-\frac{\rho}{2(1-x)}\right)\ln\frac{2x-1+\sqrt{1-\rho}}{2x-1-\sqrt{1-\rho}}-\frac{\rho}{4x(x-1)+\rho}+1-\frac{(1-2x)\rho^2}{2(1-x)(\rho-4x+4x^2)^2} &, x>1   \\[5pt]
\frac{1}{\sqrt{1-\rho}}\left(\frac{1+x^2}{1-x}-\frac{\rho}{2(1-x)}\right)\ln\frac{1+\sqrt{1-\rho}}{1-\sqrt{1-\rho}}-\frac{2x}{1-x}+\frac{1-2x}{2(1-x)} &, 0<x<1 \ \ .  \\[5pt]
\frac{1}{\sqrt{1-\rho}}\left(\frac{1+x^2}{1-x}-\frac{\rho}{2(1-x)}\right)\ln\frac{1-2x+\sqrt{1-\rho}}{1-2x-\sqrt{1-\rho}}+\frac{\rho}{4x(x-1)+\rho}-1+\frac{(1-2x)\rho^2}{2(1-x)(\rho-4x+4x^2)^2} &, x<0 
\end{cases}
\end{aligned}
\end{equation}

Taking the $\rho\ \to\ 0$ limit and isolating the logarithmic IR divergence, we have
\begin{equation}
\begin{aligned}
\tilde{C}_1(x,\rho\rightarrow0)=
\begin{cases}
\frac{1+x^2}{1-x}\ln\frac{x}{x-1}+1 &,\ x>1  \\[5pt]
\frac{1+x^2}{1-x}\ln\frac{4}{\rho}-\frac{2x}{1-x}+\frac{1-2x}{2(1-x)}&,\ 0<x<1 \ \ .\\[5pt]
\frac{1+x^2}{1-x}\ln\frac{x-1}{x}-1 &,\ x<0
\end{cases}
\end{aligned}
\end{equation}

The one-loop correction to the lightcone PDF of a single quark state is non-vanishing only when $0<x<1$, and
\begin{equation}
\begin{aligned}
q^{(1)}(x,-\mu^2/p^2)=8\zeta C_F\frac{\alpha_s}{4\pi}C_1(x,-\mu^2/p^2) \ ,
\end{aligned}
\end{equation}
where 
\begin{equation}
\begin{aligned}
C_1(x,\frac{\mu^2}{-p^2})=\frac{1+x^2}{1-x}\ln\left(\frac{\mu^2}{-x(1-x)p^2}\right)-(2-x)+\frac{1-2x}{2(1-x)} \ .
\end{aligned}
\end{equation}

The matching kernel for $\tilde{q}^{(1)}(x,\rho\rightarrow0)$ and  $q^{(1)}(x,-\mu^2/p^2)$, both  in the $\overline{\text{MS}}$ scheme, is 
\begin{equation}
\begin{aligned}
Z^{(1)}(\xi, \frac{p_z^2}{\mu^2})=&\frac{1}{4\zeta}\left[\tilde{q}^{(1)}(\xi,\rho \to 0)-q^{(1)}(\xi,-\mu^2/p^2)\right]\\[5pt]
=&2C_F\frac{\alpha_s}{4\pi}
\begin{cases}
\frac{1+\xi^2}{1-\xi}\ln\frac{\xi}{\xi-1}+1 &,\ x>1  \\[5pt]
\frac{1+\xi^2}{1-\xi}\ln\frac{4\xi(1-\xi)p_z^2}{\mu^2}-\frac{2\xi}{1-\xi}+2-\xi&,\ 0<x<1 \ \ .\\[5pt]
\frac{1+\xi^2}{1-\xi}\ln\frac{\xi-1}{\xi}-1 &,\ x<0
\end{cases}
\end{aligned}
\end{equation}
The IR divergence is cancelled between the quasi-PDF and PDF in the matching kernel as we expect that the kernel only compensates the UV difference between the quasi-PDF and PDF.

The kernel has the asymptotic behavior 
\begin{equation}
Z_1(\xi)\big|_{\xi\rightarrow\pm\infty}=-\frac{3}{2|\xi|}
\end{equation}
with $Z_{n} \equiv Z^{(n)}/(2C_F\left(\frac{\alpha_s}{4\pi}\right)^n\beta_0^{n-1})$. This leading contribution is cancelled in the RI/MOM to $\overline{\text{MS}}$ scheme matching which yields better convergence numerically \cite{Stewart:2017tvs}. 


\subsection{Two-loop result}

The two-loop correction to quasi-PDF of a single quark state can be written as
\begin{equation}
\label{A8}
\begin{aligned}
\tilde{q}^{(2)}(x,\rho, \mu^2/p_z^2)=8\zeta C_F\beta_0\left(\frac{\alpha_s}{4\pi}\right)^2\left[\tilde{C}_1(x,\rho)\ln\frac{\mu^2e^{5/3}}{4x^2p_z^2}+\tilde{C}_2(x,\rho) \ . \right]\\[5pt]
\end{aligned}
\end{equation}
The first term comes from the sub-divergence of the gluon vacuum polarization bubble. 
Taking the limit $\rho\ \to\ 0$ and isolating the logarithmic IR divergence, we have
\begin{equation}
\begin{aligned}
\tilde{C}_2(x,\rho)=
\begin{cases}
\frac{1+x^2}{1-x}\left[\ln\frac{x}{x-1}+\ln^2\frac{x}{x-1}-\text{Li}_2(\frac{1}{x})\right]+\frac{1}{1-x}&,\ x>1\\[15pt]
\frac{1+x^2}{1-x}\left[\frac{1}{2}\ln^2\frac{4}{\rho}-\text{Li}_2(x)-\frac{\pi^2}{6}-\ln(1-x)\ln\frac{4}{\rho}-\frac{1}{2}\ln^2(1-x)\right]\\[5pt]
+\left(\frac{1-2x}{2(1-x)}-x\right)\ln\frac{4}{\rho}-\frac{3}{2}\frac{1+x}{1-x}+\left(\frac{1}{2(1-x)}+\frac{2x}{1-x}\right)\ln(1-x)\\[5pt]
+\ln x^2\left[\frac{1+x^2}{1-x}\ln\frac{4}{\rho}-\frac{2x}{1-x}+\frac{1-2x}{2(1-x)}\right]&,\ 0<x<1 \ \ .\\[15pt]
\frac{1+x^2}{1-x}\left[\ln\frac{x-1}{x}-\ln^2\frac{x-1}{x}+\text{Li}_2(\frac{1}{x})\right]-\frac{1}{1-x}&,\ x<0\\[5pt]
\end{cases}
\end{aligned}
\end{equation}

The two-loop lightcone PDF is only non-vanishing for $0<x<1$ 
\begin{equation}
\begin{aligned}
q^{(2)}(x,-\mu^2/p^2)=8\zeta C_F\beta_0\left(\frac{\alpha_s}{4\pi}\right)^2\left[C_1(x,-\mu^2/p^2)\ln\left(\frac{\mu^2e^{5/3}}{-(1-x)p^2}\right)+C_2(x,-\mu^2/p^2)\right] \ ,
\end{aligned}
\end{equation}
where 
\begin{equation}
\begin{aligned}
C_2(x,\frac{\mu^2}{-p^2})=&\frac{1+x^2}{1-x}\bigg[\frac{\pi^2}{12}-\frac{1}{2}\ln^2\left(\frac{\mu^2}{-x(1-x)p^2}\right)\bigg]+2(1-x)\ln\left(\frac{\mu^2}{-x(1-x)p^2}\right)-\frac{3}{2}+x \ . \\[5pt]
\end{aligned}
\end{equation}


To this end we will define the matching kernel as
\begin{equation}
\begin{aligned}
Z^{(2)}(\xi, \frac{p_z^2}{\mu^2})=&\frac{1}{4\zeta}\left[\tilde{q}^{(2)}(\xi,\rho)
-q^{(2)}(\xi,-\mu^2/p^2)\right]\\[5pt]
=&2C_F\beta_0\left(\frac{\alpha_s}{4\pi}\right)^2\\[5pt]
&
\begin{cases}
Z_1(\xi)\ln\frac{\mu^2e^{5/3}}{4\xi^2p_z^2}+\frac{1+\xi^2}{1-\xi}\left[\ln\frac{\xi}{\xi-1}+\ln^2\frac{\xi}{\xi-1}-\text{Li}_2(\frac{1}{\xi})\right]+\frac{1}{1-\xi}\quad\ , \xi>1\\[15pt]
Z_1(\xi,\frac{p_z^2}{\mu^2})\ln\frac{\mu^2e^{5/3}}{4(1-\xi)p^2_z}+\frac{1+\xi^2}{1-\xi}\left[\frac{1}{2}\ln^2\frac{4\xi(1-\xi)p^2_z}{\mu^2}-\text{Li}_2(\xi)-\frac{\pi^2}{4}-\frac{1}{2}\ln^2(1-\xi)\right] \ \ , \\[5pt]
+2(1-\xi)\ln\frac{4\xi(1-\xi)p^2_z}{\mu^2}-\frac{3\xi}{1-\xi}-\xi+\ln(1-\xi)\quad\ , 0<\xi<1\\[15pt]
Z_1(\xi)\ln\frac{\mu^2e^{5/3}}{4\xi^2p_z^2}+\frac{1+\xi^2}{1-\xi}\left[\ln\frac{\xi-1}{\xi}-\ln^2\frac{\xi-1}{\xi}+\text{Li}_2(\frac{1}{\xi})\right]-\frac{1}{1-\xi}\quad\ , \xi<0\\[5pt]
\end{cases}
\end{aligned}
\end{equation}
which is again free from IR divergence as in the one-loop case. The IR divergence cancellation works diagram by diagram between the quasi-PDF and the PDF. However, we should have also included the convolution term $Z^{(1)}\otimes q^{(1)}$ in the kernel were we to include all the two-loop diagrams rather than just the bubble-chain diagrams. 

Again, asymptotically,  
\begin{equation}
Z_2(\xi)\big|_{\xi\rightarrow\pm\infty}\sim-\frac{9}{4|\xi|}-\frac{3}{2|\xi|}\ln\frac{\mu^2e^{5/3}}{4\xi^2p_z^2} \ . 
\end{equation}
The kernel will decay faster at large $\xi$ if we renormalize the quasi-PDF with RI/MOM as in the one-loop case.

\subsection{Three-loop result}

The three-loop quasi-PDF can be written as
\begin{equation}
\label{A14}
\begin{aligned}
\tilde{q}^{(3)}(x,\rho, \mu^2/p_z^2)=8\zeta C_F\beta_0^2\left(\frac{\alpha_s}{4\pi}\right)^3\left[\tilde{C}_1(x,\rho)\left(\ln\frac{\mu^2e^{5/3}}{4x^2p_z^2}\right)^2+2\tilde{C}_2(x,\rho)\ln\frac{\mu^2e^{5/3}}{4x^2p_z^2}+\tilde{C}_3(x,\rho)\right] \ ,\\[5pt]
\end{aligned}
\end{equation}
where the logarithmic dependence in the first two terms come from the sub-divergence of the gluon vacuum polarization bubbles.

Taking the $\rho\ \to\ 0$ limit to isolate the logarithmic IR divergence, we have 
\begin{equation}
\begin{aligned}
\tilde{C}_3(x,\rho\rightarrow0)=
\begin{cases}
2 \bigg\{\frac{1 + x^2}{1 - x}\bigg[\ln^3\frac{x}{x-1}+ 2 \ln^2\frac{x}{x -1} + \frac{\pi^2}{6}\ln\frac{x}{x-1} + \text{Li}_3\left(\frac{1}{x}\right)- 2 \zeta(3)\\[5pt]
+ 2 \left[\text{Li}_3\left(\frac{x-1}{x}\right)+ \ln\frac{x}{x - 1} \text{Li}_2\left(\frac{x-1}{x}\right)+\frac{1}{3}\ln^3\frac{x}{x - 1}+\frac{1}{2}\ln(x-1)\ln^2\frac{x}{x - 1}\right]+ \text{Li}_2\left(\frac{x - 1}{x}\right)\\[5pt]
- \frac{\pi^2}{6}+ \ln(x-1)\ln\frac{x}{x - 1}\bigg]- 2\left[\text{Li}_2\left(\frac{x - 1}{x}\right) +\frac{1}{2} \ln^2\frac{x}{x - 1} + \ln(x - 1) \ln\frac{x}{x - 1}-\frac{\pi^2}{6}\right]\\[5pt]
+ 2 \frac{1 + x}{1 - x} \ln\frac{x}{x - 1}  - \ln^2\frac{x}{x - 1} + \frac{3}{1 - x}+ \frac{\pi^2}{6}\bigg\}\quad\ ,\ x>1\\[15pt]
\left(\frac{\pi^2}{3}+\ln^2x^2\right)\left[\frac{1+x^2}{1-x}\ln\frac{4}{\rho}-\frac{2x}{1-x}+\frac{1-2x}{2(1-x)}\right]+2\ln x^2\bigg\{\frac{1+x^2}{1-x}\bigg[\frac{1}{2}\ln^2\frac{4}{\rho}-\text{Li}_2(x)\\[5pt]
-\frac{\pi^2}{6}-\ln(1-x)\ln\frac{4}{\rho}-\frac{1}{2}\ln^2(1-x)\bigg]+\left(\frac{1-2x}{2(1-x)}-x\right)\ln\frac{4}{\rho}-\frac{3}{2}\frac{1+x}{1-x}\\[5pt]
+\left(\frac{1}{2(1-x)}+\frac{2x}{1-x}\right)\ln(1-x)\bigg\}+\frac{1+x^2}{1-x} \bigg\{\frac{1}{3} \ln^3\frac{4}{\rho} - \ln(1 - x) \ln^2\frac{4}{\rho}\\[5pt]
 -\left[\frac{\pi^2}{3} - \ln^2(1 - x)\right]\ln\frac{4}{\rho}+ 2 \text{Li}_3(x)  + 4 \text{Li}_3(1-x)+4\ln(1 - x)\text{Li}_2(x)\\[5pt]
-\frac{\pi^2}{3} \ln(1 - x) + 2 \ln^2(1 - x)\ln x+ \ln^3(1 - x)\bigg\}- \frac{5 (1 + x)}{1 - x}- \frac{\pi^2}{3} \left(\frac{1 - 2 x}{2(1 - x)}- x\right)\\[5pt]
 + \frac{5+3x}{1-x} \ln(1 - x) - \frac{3+4x + 2 x^2}{2(1-x)}\ln^2(1 - x)+2\left(1+x-\frac{2x}{1-x}\right)\text{Li}_2(x)\\[5pt]
- \left[1 + 2 \left(\frac{1 - 2 x}{2 (1 - x)} - x\right) \ln(1-x)\right]\ln\frac{4}{\rho} + \left(\frac{1 - 2 x}{2 (1 - x)} - x\right) \ln^2\frac{4}{\rho}\quad\ ,\ 0<x<1 \ \ .\\[15pt]
2 \bigg\{\frac{1 + x^2}{1 - x}\bigg[\ln^3\frac{x-1}{x}- 2 \ln^2\frac{x}{x -1} + \frac{\pi^2}{6}\ln\frac{x-1}{x} - \text{Li}_3\left(\frac{1}{x}\right)+ 2 \zeta(3)\\[5pt]
- 2 \left[\text{Li}_3\left(\frac{x}{x-1}\right)+ \ln\frac{x-1}{x} \text{Li}_2\left(\frac{x}{x-1}\right)+\frac{1}{2}\ln(1-x)\ln^2\frac{x-1}{x}\right]+ \text{Li}_2\left(\frac{x}{x-1}\right)\\[5pt]
- \frac{\pi^2}{6}+\frac{1}{2} \ln^2\frac{x-1}{x }+ \ln(1-x)\ln\frac{x-1}{x}\bigg]- 2\left[\text{Li}_2\left(\frac{x}{x-1}\right)  + \ln(1-x) \ln\frac{x-1}{x}-\frac{\pi^2}{6}\right]\\[5pt]
+ 2 \frac{1 + x}{1 - x} \ln\frac{x-1}{x}  + \ln^2\frac{x-1}{x} - \frac{3}{1 - x}- \frac{\pi^2}{6}\bigg\}\quad\ ,\ x<0\\[15pt]
\end{cases}
\end{aligned}
\end{equation}

The three-loop PDF is non-vanishing only at $0<x<1$: 
\begin{equation}
\begin{aligned}
q^{(3)}(x,\frac{\mu^2}{-p^2})=8\zeta C_F\beta_0^2\left(\frac{\alpha_s}{4\pi}\right)^3\left[C_1\ln^2\left(\frac{\mu^2e^{5/3}}{-(1-x)p^2}\right)+2C_2\ln\left(\frac{\mu^2e^{5/3}}{-(1-x)p^2}\right)+C_3(x,\frac{\mu^2}{-p^2})\right] \ ,
\end{aligned}
\end{equation}
where 
\begin{equation}
\begin{aligned}
C_3(x,\frac{\mu^2}{-p^2})=&\frac{1+x^2}{1-x}\left[\frac{1}{3}\ln^3\frac{\mu^2}{-x(1-x)p^2}-\frac{\pi^2}{6}\ln\frac{\mu^2}{-x(1-x)p^2}-\frac{2}{3}\zeta(3)\right] \\[5pt]
&+2(1-x)\left(\ln\frac{\mu^2}{-x(1-x)p^2}-\ln^2\frac{\mu^2}{-x(1-x)p^2}\right)+1+\frac{\pi^2}{3}(1-x)\ . \\[5pt]
\end{aligned}
\end{equation}

Again, we define the kernel without the $Z^{(2)}\otimes q^{(1)}$ and $Z^{(1)}\otimes q^{(2)}$ terms:
%
\begin{equation}
\begin{aligned}
Z^{(3)}(\xi, p_z, \mu)=&\frac{1}{4\zeta}\left[\tilde{q}^{(3)}(\xi,p_z^2/\mu^2)
-q^{(3)}(\xi,-\mu^2/p^2)\right]\\[5pt]
=&2C_F\beta_0^2\left(\frac{\alpha_s}{4\pi}\right)^3\\[5pt]
&
\begin{cases}
Z_1(\xi)\ln^2\frac{\mu^2e^{5/3}}{4\xi^2p^2_z}+2Z_2(\xi)\ln\frac{\mu^2e^{5/3}}{4\xi^2p^2_z}\\[5pt]
+2\frac{1 + \xi^2}{1 - \xi}\bigg\{\ln^3\frac{\xi}{\xi-1}+ 2 \ln^2\frac{\xi}{\xi -1} + \frac{\pi^2}{6}\ln\frac{\xi}{\xi-1}+ \text{Li}_3\left(\frac{1}{\xi}\right)\\[5pt] 
+ 2 \left[\text{Li}_3\left(\frac{\xi-1}{\xi}\right)+ \ln\frac{\xi}{\xi - 1} \text{Li}_2\left(\frac{\xi-1}{\xi}\right)+\frac{1}{3}\ln^3\frac{\xi}{\xi - 1}+\frac{1}{2}\ln(\xi-1)\ln^2\frac{\xi}{\xi - 1}\right]\\[5pt]
+ \text{Li}_2\left(\frac{\xi - 1}{\xi}\right)- \frac{\pi^2}{6}+ \ln(\xi-1)\ln\frac{\xi}{\xi - 1}- 2 \zeta(3)\bigg\}\\[5pt]
- 4\bigg[\text{Li}_2\left(\frac{\xi - 1}{\xi}\right) +\frac{1}{2} \ln^2\frac{\xi}{\xi - 1}+ \ln(\xi - 1) \ln\frac{\xi}{\xi - 1}-\frac{\pi^2}{6}\bigg]\\[5pt]
+2\left[2 \frac{1 + \xi}{1 - \xi} \ln\frac{\xi}{\xi - 1}  - \ln^2\frac{\xi}{\xi - 1} + \frac{3}{1 - \xi}+ \frac{\pi^2}{6}\right],& \xi>1\\[15pt]
-Z_1(\xi,\frac{p_z^2}{\mu^2})\ln^2\frac{\mu^2e^{5/3}}{4(1-\xi)p^2_z}+2Z_2(\xi,\frac{p_z^2}{\mu^2})\ln\frac{\mu^2e^{5/3}}{4(1-\xi)p^2_z}\\[5pt]
+\frac{1+\xi^2}{1-\xi}\bigg\{\frac{1}{3}\ln^3\frac{4\xi(1-\xi)p^2_z}{\mu^2}-\frac{\pi^2}{6}\ln\frac{4\xi(1-\xi)p^2_z}{\mu^2}+2\text{Li}_3(\xi)+4\text{Li}_3(1-\xi)\\[5pt]
+2\ln(1-\xi)\text{Li}_2(\xi)+2\ln^2(1-\xi)\ln \xi-\frac{2\pi^2}{3}\ln(1-\xi)+\frac{2}{3}\zeta(3)\bigg\}\\[5pt]
+2(1-\xi)\left[\ln^2\frac{4\xi(1-\xi)p^2_z}{\mu^2}+\ln\frac{4\xi(1-\xi)p^2_z}{\mu^2}\right]+2\left(1+x-\frac{2x}{1-x}\right)\text{Li}_2(x)\\[5pt]
-\frac{5(1+\xi)}{1-\xi}-1-\frac{\pi^2}{3}\left(1-2x+\frac{2x}{1-x}\right)+\frac{2\ln(1-x)}{1-x}-\frac{x(1+x)}{1-x}\ln^2(1-x),& 0<\xi<1 \ ,\\[15pt]
Z_1(\xi)\ln^2\frac{\mu^2e^{5/3}}{4\xi^2p^2_z}+2Z_2(\xi)\ln\frac{\mu^2e^{5/3}}{4\xi^2p^2_z}\\[5pt]
+2 \frac{1 + \xi^2}{1 - \xi}\bigg\{\ln^3\frac{\xi-1}{\xi}- 2 \ln^2\frac{\xi}{\xi -1} + \frac{\pi^2}{6}\ln\frac{\xi-1}{\xi}- \text{Li}_3\left(\frac{1}{\xi}\right)\\[5pt]
- 2 \left[\text{Li}_3\left(\frac{\xi}{\xi-1}\right)+ \ln\frac{\xi-1}{\xi} \text{Li}_2\left(\frac{\xi}{\xi-1}\right)+\frac{1}{2}\ln(1-\xi)\ln^2\frac{\xi-1}{\xi}\right]\\[5pt]
+ \text{Li}_2\left(\frac{\xi}{\xi-1}\right)- \frac{\pi^2}{6}+\frac{1}{2} \ln^2\frac{\xi-1}{\xi }+ \ln(1-\xi)\ln\frac{\xi-1}{\xi}+ 2 \zeta(3)\bigg\}\\[5pt]
- 4\bigg[\text{Li}_2\left(\frac{\xi}{\xi-1}\right)  + \ln(1-\xi) \ln\frac{\xi-1}{\xi}-\frac{\pi^2}{6}\bigg]\\[5pt]
+ 2\left[2 \frac{1 + \xi}{1 - \xi} \ln\frac{\xi-1}{\xi}  + \ln^2\frac{\xi-1}{\xi} - \frac{3}{1 - \xi}- \frac{\pi^2}{6}\right],& \xi<0\\[15pt]
\end{cases}
\end{aligned}
\end{equation}
where IR divergence is cancelled. And the asymptotic behavior 
%
%
\begin{equation}
Z_3(\xi)\big|_{\xi\rightarrow\pm\infty}\sim-\frac{27+2\pi^2}{4|\xi|}-\frac{9}{2\xi}\ln\frac{\mu^2e^{5/3}}{4\xi^2p^2_z}-\frac{3}{2|\xi|}\ln^2\frac{\mu^2e^{5/3}}{4\xi^2p^2_z}
\end{equation}
will again decay faster if RI/MOM renormalization is applied to the quasi-PDF.

\bibliographystyle{apsrev4-1}
\bibliography{refs}

\end{document}